\documentclass[prl,preprint,showpacs,aps]{revtex4}

\usepackage[dvips]{graphics}
\usepackage{epsfig}
\usepackage{dcolumn}
\usepackage[T1]{fontenc}
\usepackage[latin1]{inputenc}
\usepackage{latexsym}
\usepackage{natbib}
\usepackage{amsmath}
\usepackage{amssymb}
\usepackage{amsfonts}

\begin{document}

\title[bold]{Laser-induced hydrodynamic instability of fluid interfaces}

\author{Alexis Casner}
\email{acasner@cribx1.u-bordeaux.fr}
\author{Jean-Pierre Delville}
\email{delville@cribx1.u-bordeaux.fr}

\affiliation{Centre de Physique Mol\'{e}culaire Optique et Hertzienne,UMR CNRS/Universit\'{e} 5798, Universit\'{e}
Bordeaux I, 351 Cours de la Libération, F-33405 Talence cedex, France}

\date{\today}
\begin{abstract}
We report on a new class of electromagnetically-driven fluid interface instability. Using the optical radiation
pressure of a cw laser to bend a very soft near-critical liquid-liquid interface, we show that it becomes unstable
for sufficiently large beam power $P$, leading to the formation of a stationary beam-centered liquid micro-jet. We
explore the behavior of the instability onset by tuning the interface softness with temperature and varying the size
of the exciting beam. The instability mechanism is experimentally demonstrated. It simply relies on total reflection
of light at the deformed interface whose condition provides the universal scaling relation for the onset $P_{S}$ of
the instability.
\end{abstract}


\maketitle
 Deformations and break-up of fluid interfaces under an applied field play a significant role in
scientific and technologic endeavors. Since the early works of Zeleny \cite{zeleny} and further investigations of
Taylor \cite{taylorjets}, interface instabilities driven by electric fields are the most familiar examples. They
represent nowadays the corner stone of numerous industrial processes as different as electro-spraying
\cite{ganancalvo}, electrospinning of polymer fibers \cite{electrospinning} or surface relief patterning
\cite{schaffer}. From the fundamental point of view they also illustrate a simple and fascinating example of
behavior leading to finite time singularity \cite{nagelspout}. A uniform magnetic field can as well destabilize
fluid interfaces and form well-organized peak structures \cite{mahr} or elongate magnetic droplets
\cite{bacrisalin}, though no liquid ejection occurs in that case. These deformations, as well as those driven by the
acoustic radiation pressure \cite{elrod}, were essentially used to characterize in a non contact manner, the
mechanical properties of fluid interfaces \cite{cinbis,bacri2}.

Recently, with the increasing importance of optical forces in emerging nano/bio-technologies, this strategy has been
successfully extended to the optical deformation of soft materials or micro-objets and brought new insights into
their viscoelastic properties \cite{vesicules,kasbiophys,sakai02}. However, as the bending of fluid surfaces by the
optical radiation pressure is usually weak, further developments require larger beam intensities with smaller
disturbing effects. Interface disruption under the applied field could constitute a major drawback for this
promising method and should therefore be investigated. Yet, even if non linear elongations of red blood cells
\cite{kasbiophys} and droplets disruption \cite{zhang} were already shown, to the best of our knowledge interface
instability driven by the optical radiation pressure has never been observed nor discussed in detail.

Using very soft transparent liquid interfaces to enhance optical radiation pressure effects, we show in the present
Letter that, contrary to expectations, the hump induced by a continuous laser wave becomes unstable for sufficiently
large beam powers. To analyze this new opto-hydrodynamic instability, we explore the universal behavior of its onset
on near-critical liquid-liquid interfaces by varying the interface softness with a temperature scanning, and by
changing the size of the exciting beam. When properly rescaled, the dispersion in measured onsets reduces to a
single master behavior which is retrieved theoretically from simple arguments. In fact we demonstrate that the total
reflection condition of light at the interface defines an onset in power which agrees with the observed scaling law.
Finally we show that the optically-driven interface instability leads to the formation of a stationary beam-centered
liquid micro-jet emitting droplets, which anticipates the bases for new applications in microfluidics and liquid
micro-spraying.

\textit{Experiments}.---Opto-hydrodynamic instabilities of fluid interfaces are realized in a quaternary liquid
mixture made of toluene, sodium dodecyl sulfate (SDS), n-butanol and water. At room temperature we prepare a
water-in-oil micellar phase of microemulsion whose composition in weight fraction is 70 \% toluene, 4 \% SDS, 17\%
n-butanol and 9\% water. Close to $T_{C}$, where $T_{C}= 35°C$ is the liquid-liquid critical temperature, the
mixture belongs to the universality class (d=3,n=1)  of the Ising model \cite{eric}. For a temperature $T >T_{C}$ it
separates in two micellar phases of different concentrations $\Phi_{1}$ and $\Phi_{2}$. The use of these mixtures
was motivated by the fact that significant interface deformations by optical radiation, without nonlinear
propagation effect or disturbing thermal coupling \cite{eric}, require weak surface tension and buoyancy. Indeed
both effects vanishes when the critical point is neared respectively with the critical exponents $2 \nu = 1.26 $ and
$\beta = 0.325$. Then the surface tension $\sigma$ of phase-separated supramolecular liquids is intrinsically small.
For example from $\sigma=\sigma_{0}(\frac{T}{T_{C}}-1)^{2\nu}$, where $\sigma_{0} = 1.04 \; 10^{-4} N/m $ in our
case, we get $\sigma \approx 10^{-7} N/m $ at $(T-T_{C})= 1.5 K$ a value typically $10^{6}$ times smaller than that
of the water free surface at room temperature.

\begin{figure}
\begin{center}
\includegraphics*[width=8cm,keepaspectratio]{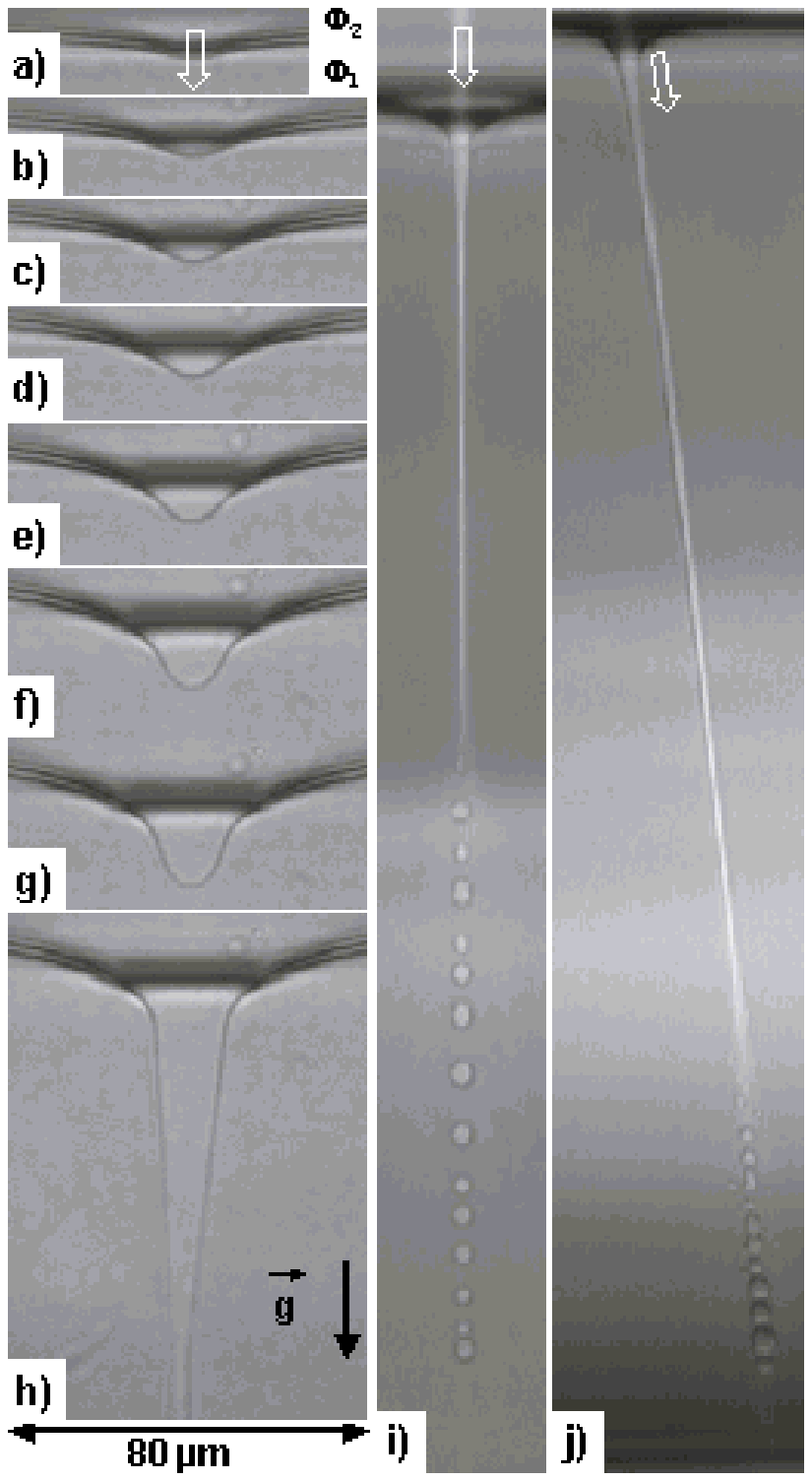}
\end{center}
\caption{(a-h) Variation of the optical interface bending versus beam power until instability when $(T-T_{C})= 15 K$
and $\omega_{0} = 3.5 µm $ : (a) $P$ = 310 mW, (b) $P$ = 500 mW, (c) $P$ = 620 mW, (d) $P$ = 740 mW, (e) $P$ = 920
mW, (f) $P$ = 1110 mW, (g) $P$ = 1170 mW , and (h) $P =P_{S} = 1230 \; mW $. For $P =P_{S}$ the interface becomes
unstable leading to the formation of a stationary liquid jet similar to that illustrated in the overview (i) for
$(T-T_{C})= 6 K$, $\omega_{0} = 3.5 µm $ and $P =P_{S} = 490 \; mW $. (j) Inclined liquid jet obtained for the same
experimental conditions as (i) but for $P = 770 \; mW > P_{S} $. The laser beam is represented by the arrows, and
the height of (i) and (j) is 1mm.} \label{fig1}
\end{figure}
The experimental configuration is schematically presented in Fig. 1a. The mixture is enclosed in a fused-quartz cell
of path length l=2 mm   that is thermally-controlled at the temperature $T$ above $T_{C}$ (temperature accuracy:
$\Delta T \leqslant 0.05 K$). Since density (resp. index of refraction) of water is larger (resp. smaller) than that
of toluene, the micellar phase of larger concentration $\Phi_{1}$ is located below the low micellar concentration
phase $\Phi_{2}$, while its refractive index $n_{1}$ is smaller than $n_{2}$ of $\Phi_{2}$. The difference in
density is given by $\rho_{1}-\rho_{2}=(\Delta \rho)_{0}(\frac{T}{T_{C}}-1)^{\beta}$ with $(\Delta \rho)_{0}\approx
285 \; kg.m^{-3}$ , and the Clausius-Mossotti relation close to the critical point leads to
$n_{2}-n_{1}=(\frac{\partial n}{\partial \rho})_{T} (\rho_{2}-\rho_{1})$  with $(\frac{\partial n}{\partial
\rho})_{T} \approx -1.22 \; 10^{-4} \; m^{3}.kg^{-1}$. The bending of the liquid-liquid meniscus is driven by a
linearly polarized $ TEM_{00}$ cw $Ar^{+}$ laser (wavelength in vacuum, $\lambda_{0}= 5145 $ \AA ) propagating
vertically downward from $\Phi_{2}$ to $\Phi_{1}$. The beam is focused on the interface by a 10X microscope lens to
ensure a cylindrical symmetry of the intensity profile around the meniscus. We adjust also the beam-waist
$\omega_{0}$ by changing the distance between a first lens (focal length f=1 m) and the 10X objective.

 As already demonstrated, the optical radiation pressure originates from light momentum discontinuity at the
interface. It always pushes normally to this interface the fluid with larger refractive index into that with smaller
refractive index, regardless of the direction of propagation \cite{Ash2}. As a result, radiation pressure acts
downwards in our case (see Fig. 1), and the height $h(r)$ of the induced deformation is determined by a balance with
the buoyancy and Laplace restoring forces. Then, at steady state $h(r)$ is described by \cite{sakai02}:

\begin{equation}
  (\rho_{1}-\rho_{2}) g h(r)  - \sigma  \frac{1}{r} \frac{d}{d r} \Big (\frac{r h^{'}(r)}{\sqrt{1 + h^{'}(r)^{2}}}
   \Big ) = \Pi(r)
\label{equa1}
\end{equation}
Primes denote derivatives versus $r$ and $g$ is the gravity constant. In the general case, the optical radiation
pressure $\Pi(r)$ is given by \cite{borzdov}:
\begin{equation}
 \Pi(r) = \frac{n_{2} \cos^{2} \theta_{2}}{c} \Big [1+R-\frac{\tan \theta_{2}}{\tan \theta_{1}}T \Big ] I(r)
\end{equation}
where $ I(r) = \frac{2P}{\pi\omega_{0}^{2}}\exp(\frac{-2r^{2}}{\omega_{0}^{2}})$ is the intensity of the incident
wave, $P$ the beam power and c is the light velocity in vacuum. $R$ and $T$ are the Fresnel coefficients of
reflection and transmission in energy. $\theta_{2}$ and $\theta_{1}$ are respectively the incident and the
refraction angles versus the interface normal and are given by $\cos \theta_{2}= \frac{1}{\sqrt{1 + h^{'}(r)^{^2}}}$
and the Snell relation.
\begin{figure}
\begin{center}
\includegraphics*[width=7cm,keepaspectratio]{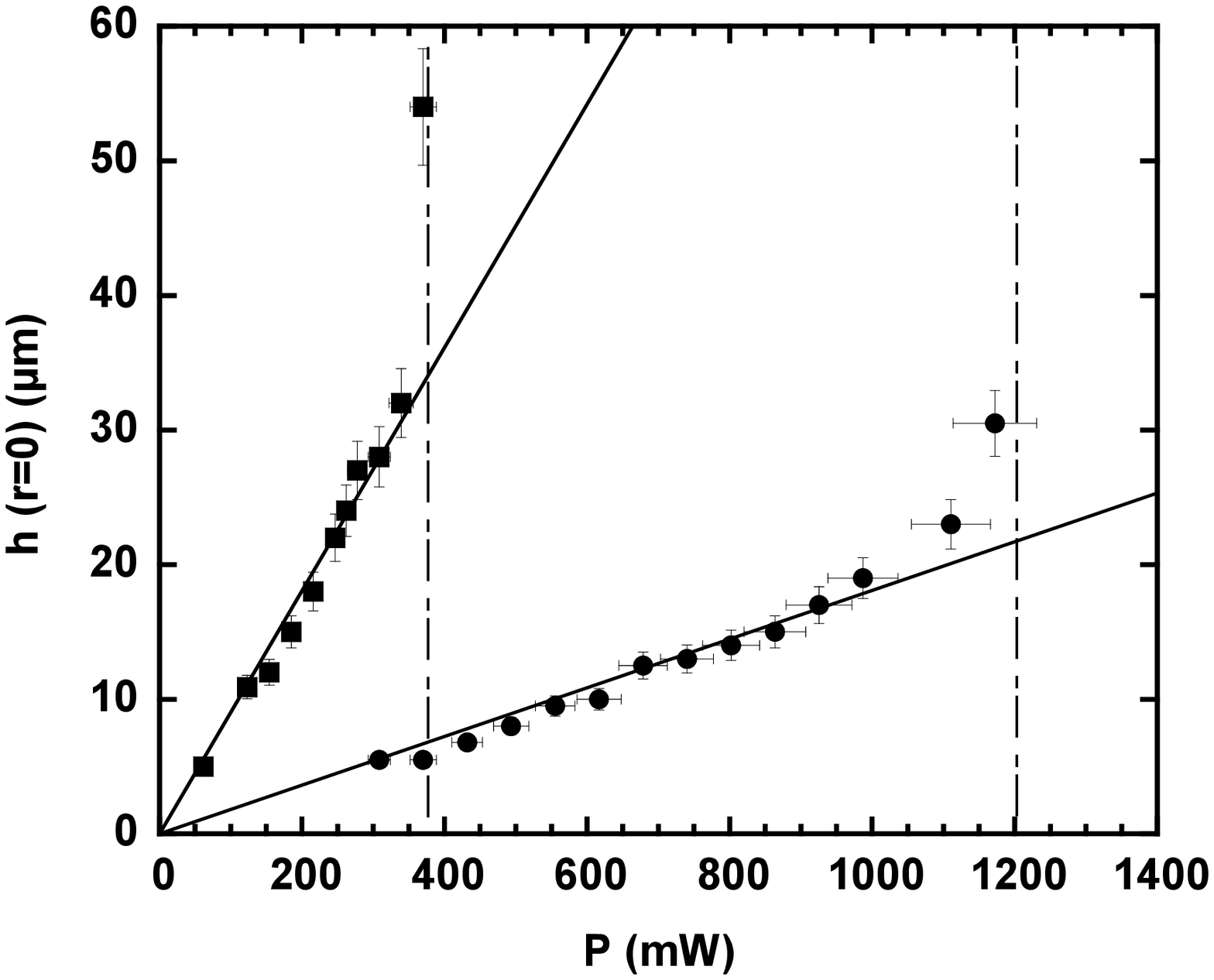}
\end{center}
\caption{Evolution of the height of the deformation $h(r=0)$ versus $P$ for the extreme values of the Bond number
range investigated. Parameters are respectively: ($\bullet$) $T-T_{C}=15 K$ and $\omega_{0}=3.5 µm$ corresponding to
the profiles represented in Figs. (1a-1h) and ($\blacksquare$) $T-T_{C}=1.5 K$ and $\omega_{0}=11 µm$. Full lines
are linear fits of the weak deformation regime. Broken lines indicate the onset $P_{S}$ above which the interface is
unstable.} \label{fig2}
\end{figure}

\textit{Results}.--- For small bending curvatures, i.e. for $h^{'}(r) \ll 1 $, deformations are function of an
optical Bond number $Bo$ defined as the ratio of buoyancy over the Laplace force \cite{prlgiant}:  $Bo =
(\frac{\omega_{0}}{l_{C}})^{2}$ , where $l_{C} = \sqrt{\frac{\sigma}{(\rho_{1}-\rho_{2})g}} $  is the capillary
length. In the present analysis, we used values of $\omega_{0}$  and $(T-T_{C})$ ranging respectively from 3.5 to 11
$µm$ and 1.5 to 15K in order to work in the $Bo < 1 $ regime (we have experimentally 0.006 < $Bo$ < 0.54) and
reproduce the case of classical liquid free surfaces. A typical evolution of the induced deformations versus
incident beam power $P$ is illustrated in Fig. 1a-1h for $T-T_{C} = 15 K$ and $\omega_{0} = 3.5 µm$, i.e. for the
smallest $Bo$ investigated. Fig. 2 depicts also the variation of $h(r=0)$ versus $P$ for the two extreme values of
our $Bo$ number range; experimental errors are $\frac{\Delta P}{P} \approx \frac{\Delta \omega_{0}}{\omega_{0}} \leq
5 \% $ . As expected, $h(r=0)$ is linear in $P$ as far as $h^{'}(r) \ll 1 $, i.e. typically for $ P \leq 970 \; mW $
in the example presented (Fig. 1a-1e). The measured slope is moreover in complete accordance with the linear
behavior predicted \cite{prlgiant}. Then, with further increase in $P$, $h(r=0)$ rapidly deviates from the linear
regime ($970 \leq P \leq 1170 \; mW $) and suddenly diverges when $P$ reaches a power onset $P_{S}=1230 \; mW $ (see
the transition from Fig. 1g to Fig. 1h). Fig. 2 shows that this behavior is observed over a large range of the
parameter space and that instability occurs even more rapidly ($ P_{S}= 370 \; mW $) when the critical point is
neared ($T-T_{C} = 1.5 K$ ), despite an increase in beam-waist ($\omega_{0}=11 µm$).

 Consequently, since $P_{S}$ vary noticeably in the range of $Bo$ investigated, we analyzed its behavior over large variations
in $\omega_{0}$ and ($T-T_{C}$). At first, as illustrated in Fig. 3 over one order of magnitude in ($T-T_{C}$),
$P_{S}$ varies linearly versus $\omega_{0}$: the instability onset decreases with focusing. Moreover results show
that the slope of $P_{S}$ versus $\omega_{0}$ also decreases when the critical point is neared. This behavior is
much better illustrated in the Inset of Fig. 3 for the smallest beam-waist used $\omega_{0} = 3.5 µm$. A power law
fit leads to $P_{S} \propto (T-T_{C})^{1.01 \pm 0.05}$.
\begin{figure}[t]
\begin{center}
\includegraphics*[width=7cm,keepaspectratio]{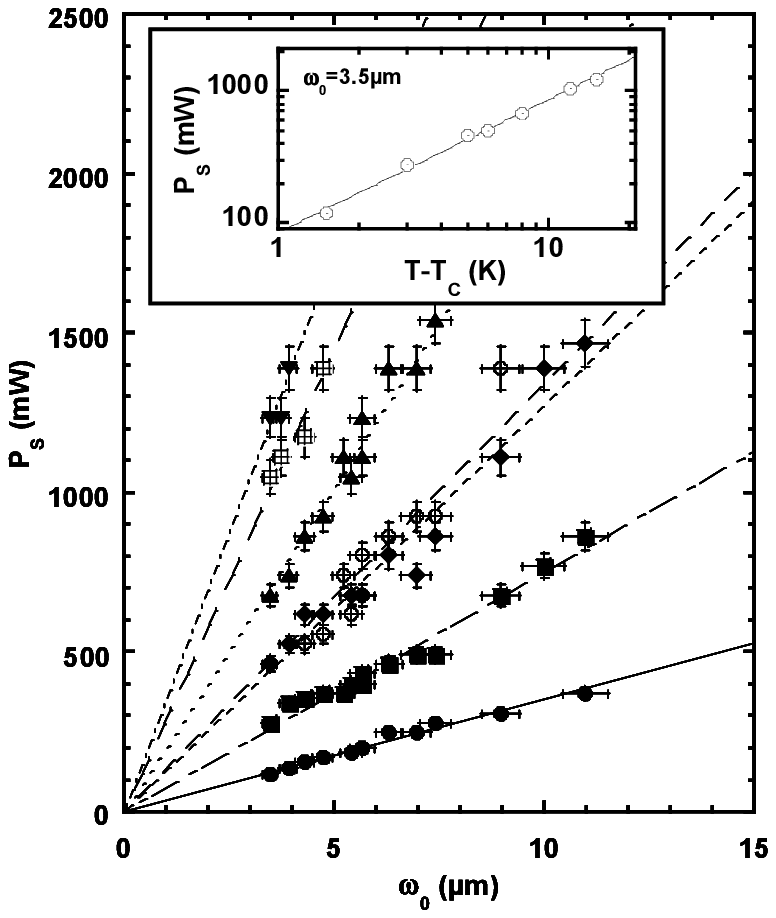}
\end{center}
\caption{Variation of the instability onset $P_{S}$ versus beam waist $\omega_{0}$ as a function of $T-T_{C}$.
Parameters are: $(T-T_{C})= 1.5 K ( \bullet ), 3 K ( \blacksquare ), 5 K ( \blacklozenge ), 6 K (\circ)$, $ 8 K (
\blacktriangle ), 12 K (\boxplus )$, and $ 15 K ( \blacktriangledown )$. Lines represent linear fits. Inset: Log-log
plot of the variation of $P_{S}$ versus $T-T_{C}$  for $\omega_{0}= 3.5 µm $. The full line is a power-law fit
giving $ P_{S} \propto (T-T_{C})^{1.01 \pm 0.05} $.}
 \label{fig3}
\end{figure}

\textit{Discussion}.--- While interface deformations do not depend on the direction of propagation at low laser
excitations, this is no more true at higher field strengths. Instead of a liquid jet, huge tethered deformations are
induced by an upward propagating beam, but they remain always \textit{stable}, despite surprising aspect ratios
\cite{coloq7}. A simple physical mechanism that differentiates both cases is the possibility to reach total-internal
reflection when the beam propagates from $\Phi_{2}$ to $\Phi_{1}$, i.e. from the large to the low refractive index
phase. Furthermore, as it can be seen from Eq. (2), the transfer of momentum at the interface is \textit{maximum}
under total reflection condition because $R=1$ and $T=0$ in that case. A comparison with the momentum transfer at
normal incidence shows that the radiation pressure is thus \textit{enhanced} by a factor
$(1+\frac{n_{2}}{n_{1}})^{2} \simeq 4 $ ($n_{1} \simeq n_{2}$ in our case). This factor is obviously underestimated
because secondary reflections and even self-guiding of light inside the deformation will also contribute to a
further increase. Then, let us express the total reflection condition:
\begin{equation}\label{equa3}
  \sin \theta_{2} = \frac{h^{'}(r)}{\sqrt{1+h^{'}(r)^{2}}} > \frac{n_{1}}{n_{2}}
\end{equation}
Since experiments are realized in the $Bo$ < 1  regime, buoyancy does not play any important role and can thus be
neglected. On the other hand, we expand the expression of the radiation pressure given by Eq. (2) to first order in
$n_{2}-n_{1}$ because $\Phi_{2}$ and $\Phi_{1}$ are coexisting phases of close composition due to the vicinity of
the critical point. Eq. (1) becomes:
\begin{equation}\label{equa4}
  \sigma  \frac{1}{r} \frac{d}{d r} \Big (\frac{r h^{'}(r)}{\sqrt{1 + h^{'}(r)^{2}}} \Big ) = \frac{2 n_{2}}{c} \Big (
  \frac{n_{2}-n_{1}}{n_{2}+n_{1}} \Big ) \frac{2P}{\pi \omega_{0}^{2}} \exp \Big (-\frac{2r^{2}}{\omega_{0}^{2}}
  \Big)
\end{equation}
Integration of Eq.  (4) gives directly the expression of $\sin \theta_{2}$  and Eq. (3) leads to:
\begin{equation}\label{equa5}
  \frac{n_{2}}{c} \Big ( \frac{n_{2}-n_{1}}{n_{2}+n_{1}} \Big ) \frac{P}{\pi \sigma}
  \frac{1}{r} \Big [ 1-\exp(\frac{-2r^{2}}{\omega_{0}^{2}}) \Big ] > \frac{n_{1}}{n_{2}}
\end{equation}
Total reflection will then be locally reached for $\frac{\sqrt{2} r}{\omega_{0}} = 1.121 $, which renders the left
hand term of Eq. (5) maximum. This condition defines an onset in \textit{power}
\begin{equation}\label{equa6}
  P > \frac{1.121 \; \pi}{0.715 \; \sqrt{2}} \frac{n_{1}}{n_{2}} \Big (1 + \frac{n_{1}}{n_{2}} \Big ) \frac{\sigma c}{n_{2}-n_{1}}
  \; \omega_{0}
\end{equation}
\begin{figure}[t]
\begin{center}
\includegraphics*[width=7cm,keepaspectratio]{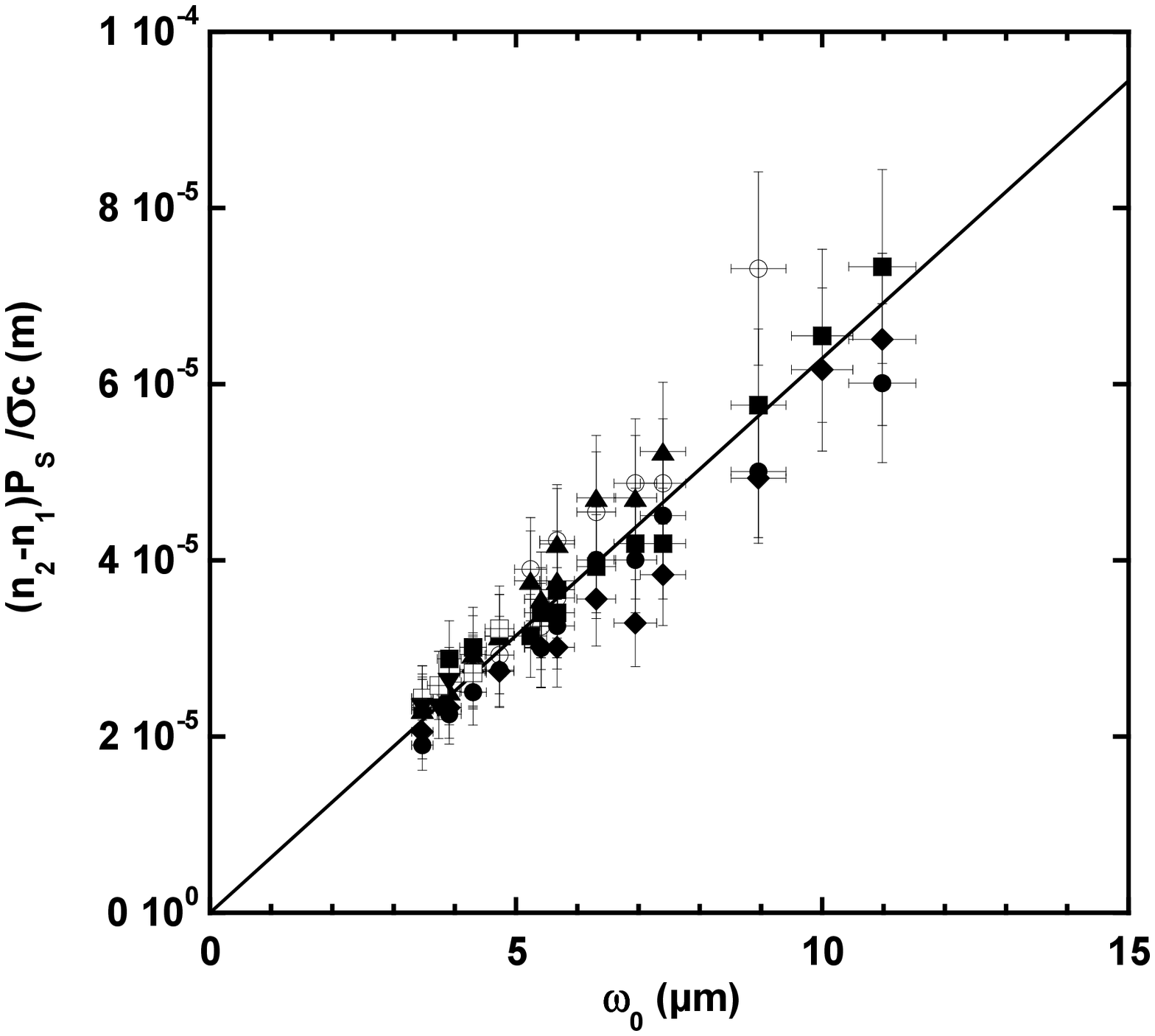}
\end{center}
\caption{Rescaling of the set of dispersed data presented in Fig. 3 according to the behavior predicted by Eq. (6).
Symbols are the same. The full line represents a linear fit.} \label{fig4}
\end{figure}
which is linear in $\omega_{0}$ and behaves as $(T-T_{C})^{2 \nu - \beta}$ , i.e. with the exponent $2 \nu - \beta =
0.935$, in very close agreement with the critical behavior of $P_{S}$. In looking for scaling, Eq. (6) also provides
a natural way for presenting the instability onsets. This is illustrated on Fig. 4 where the whole set of data is
brought onto a single master curve. As expected, a linear behavior of $ P_{S}$ versus $\frac{\sigma c
\omega_{0}}{n_{2}-n_{1}}$ is observed and a linear fit leads to $P_{S}(W) = (6.3 \pm 0.3) 10^{-6} [\frac{\sigma
c}{n_{2}-n_{1}}] \; \omega_{0} (µm)$ while the predicted slope is $ 6.9 \; 10^{-6} $. Optohydrodynamic interface
instability is then clearly triggered by the total reflection of the exciting beam inside the induced deformation.
Hence, the brightness of the liquid jets (see Fig. 1i,j) confirms that they behave as liquid waveguides.

 Finally, since matter with large refractive index is always trapped in high beam intensity regions by optical forces
\cite{Ash1}, the optical destabilization of liquid interfaces leads to the formation of a stationary beam-centered
jet which eventually breaks into a spray of nearly monodisperse droplets (Fig. 1i). Bearing in mind that our system
has ultralow surface tension and that the density difference between the phases is also weak, it can be calculated
that the radius of the induced jet is much smaller than the typical size for which fluctuation effects become
important for pinching \cite{brennershinagel,eggersrmp,eggersnanojet}. The break-up mechanism, which deserves
further investigation, is very robust and convenient because inclined jets can also be easily created by tilting the
laser beam, as shown on Fig. 1j.

    In conclusion, we have presented a new electromagnetic instability mechanism of fluid interfaces driven by the optical
radiation pressure of a cw laser wave. It opens the way to optical streaming of fluid interfaces, even if laser
pulses could be necessary for stiffer interfaces \cite{zhang}. The stability of the resulting jet, as well as the
regularity of the produced micro-droplets, offers also new perspectives for optical guiding of microfluidic flows.

\begin{acknowledgments}
This work was partly supported by the CNRS and the Conseil Régional d'Aquitaine.
\end{acknowledgments}

\end{document}